\documentclass[pra,twocolumn,showpacs]{revtex4-1}
\usepackage{hyperref}
\usepackage{color}
\usepackage{times}
\usepackage{amsmath}
\usepackage{amssymb}
\usepackage{amsfonts}
\usepackage{bm}
\usepackage{graphicx} 
\usepackage{multirow}

 \newcommand{\eref}[1]{(\ref{#1})} 
\newcommand{\fref}[1]{Fig.~\ref{#1}}

\newcommand{\mint}[1]{\int d^3 #1\;}

\newcommand{\x}{\mathbf{x}}
\newcommand{\rr}{\mathbf{r}}

\newcommand{\kk}{\mathbf{k}}

\newcommand{\q}{\mathbf{q}}
\newcommand{\eq}[2]{\begin{equation}\label{#1}#2 \end{equation}}
\newcommand{\eqn}[1]{\begin{eqnarray}#1 \end{eqnarray}}

\newcommand{\ec}{\epsilon_{\rm{cut}}}

\newcommand{\rC}{\textbf{C}} 
\newcommand{\rI}{\textbf{I}} 

\newcommand{\PC}{\mathcal{P}}

\newcommand{\intV}[1]{\int d^3 #1\;}

\newcommand{\cf}{\psi} 
\def\x{\mathbf{r}}

\newcommand{\ecut}{\epsilon_{\rm cut}}

\definecolor{MyGray}{rgb}{.2,.5,.3}

\begin{document}

\title{The stochastic projected Gross-Pitaevskii equation}
\author{S. J. Rooney} 
\author{P. B. Blakie} 
\author{A.~S. Bradley} 
\address{Jack Dodd Centre for Quantum Technology, Department of Physics, University of Otago, Dunedin, New Zealand.}
\date{\today}
\begin{abstract}
We have achieved the first full implementation of the stochastic projected Gross-Pitaevskii equation for a three-dimensional trapped Bose gas at finite temperature. Our work advances previous applications of this theory, which have only included growth processes, by implementing number-conserving scattering processes. 
We evaluate an analytic expression for the coefficient of the scattering term and compare it to that of the growth term in the experimental regime, showing the two coefficients are comparable in size. 
We give an overview of the numerical implementation of the deterministic and stochastic terms for the scattering process, and use simulations of a condensate excited into a large amplitude breathing mode oscillation to characterize the importance of scattering and growth processes in an experimentally accessible regime. We find that in such non-equilibrium regimes the scattering can dominate over the growth, leading to qualitatively different system dynamics. In particular, the scattering causes the system to rapidly reach thermal equilibrium without greatly depleting the condensate, suggesting that it provides a highly coherent energy transfer mechanism. 
\end{abstract}
\maketitle
\section{Introduction}
Ultracold atomic gases provide a unique environment to observe many body quantum phenomena on a mesoscopic scale, allowing microscopically derived field theories to be readily compared with experiments \cite{Bloch2008a,Dalfovo1999}.  
For a Bose gas at zero temperature a Bose-Einstein condensate (BEC) forms in which essentially all the atoms occupy a single spatial mode whose equilibrium and dynamics is well-described by the Gross-Pitaevskii equation (GPE) \cite{Dalfovo1999,Denschlag2000a}.  
A growing number of experiments have been performed in the \textit{finite temperature regime}, i.e.~at temperatures of order the critical temperature where many atoms are thermally excited out of the condensate, e.g.~studies of collective modes \cite{Jin1997}, vortex nucleation and decay \cite{Raman2001,Haljan2001,Weiler08a}, condensate growth~\cite{Miesner1998,QKPRLIII}, phase transition dynamics \cite{Weiler08a}, and superfluid turbulence \cite{Neely2012a}.  A quantitative description applicable to simulating many of these experiments, where thermal fluctuations and dynamics are important, remains a major technical challenge \cite{Proukakis08a,Blakie08a}.

A promising direction of investigation for the finite temperature regime has been the generalization of Gross-Pitaevskii theory to describe the entire low energy part of the system. One approach, typically referred to as the \textit{classical field method} \cite{Davis2001b,Goral2002}, involves propagating the GPE with many suitably randomized modes. An alternative approach, central to the focus of this paper, is to extend the GPE with noise and damping terms that represent the coupling to a thermal reservoir of high energy atoms. Recently such \textit{stochastic} GPEs have been applied to a broad range of problems, including defect formation across phase transitions \cite{Bradley08a,Weiler08a,Damski10a,Das2012a}, the decay of vortices~\cite{Rooney10a,Rooney11a,Rooney2012a} and solitons~\cite{Cockburn10a}, and polariton \cite{Wouters09a} and spinor \cite{Su11a,Su12a} condensates. The technique has seen quite extensive applications to low dimensional systems where thermal fluctuations can prevent a true condensate from forming \cite{Stoof2001,Proukakis06c,Proukakis06b,Proukakis06a,Cockburn09a,Cockburn10a,Cockburn11b,Cockburn11c,Cockburn11a,Davis12a,Gallucci12a,Cockburn12a}.

While phenomenological arguments can be used to obtain a generic stochastic GPE, it is possible to derive such a description from the microscopic theory of a Bose gas. Such formal derivations have been 
carried out by the groups of Stoof \cite{Stoof1997,Stoof1999,Stoof2001} and Gardiner \cite{SGPEI, SGPEII, SGPEIII} validating this approach as an \textit{ab initio} description of non-equilibrium dynamics.

In this work we present the first complete implementation of the stochastic projected Gross-Pitaevskii equation (SPGPE), first introduced by Gardiner and Davis in 2003  \cite{SGPEII}. The basic idea of the SPGPE is to sub-divide the system modes into two regions: (i) a high energy region, referred to as the \textit{incoherent} $\rI$-region, consisting of sparingly occupied modes; (ii) a low energy region, the  $\rC$-region, of appreciably occupied modes, i.e~not only the condensate, but all other highly Bose-degenerate modes. The use of projection operators to define these regions is a key aspect in the derivation of the SPGPE theory, and lead to an explicit projection operator in the equation of motion for the $\rC$-region. The thermal effects upon the $\rC$-region are described by two distinct processes:   (i) \emph{growth processes} where collisions between two $\rI$-region atoms leads to a change in population of the $\rC$-region;  (ii) \emph{scattering processes} corresponding to collision between atoms in the $\rC$- and $\rI$-regions in which energy is transferred but particles are conserved.
Simulations including the growth process are relatively straightforward to carry out as the noise is additive and uncorrelated, and the damping term is proportional to the Gross-Pitaevskii evolution operator.  Implementing the scattering process is more technically challenging, as the noise is multiplicative and spatially correlated, and the associated damping term involves an intricate calculation of current in the $\rC$-region. To date all simulations of the SPGPE have only included growth processes \footnote{More generally, all stochastic GPE calculations to date have been made with equations roughly equivalent to including growth processes.}. The neglect of the scattering  process has been argued as a reasonable approximation for systems near equilibrium, but is expected to play an important role in non-equilibrium regimes \cite{Blakie08a}. Indeed,  work on condensate growth within quantum kinetic theory \cite{QKPRLI,QKPRLII,QKPRLIII} has shown that an analogous scattering reservoir interaction has a large effect on condensation dynamics \cite{QKPRLII,QKVI}.   

 The broad outline of our paper is as follows: 

In Sec.~\ref{Sec:SPGPEformalism} we briefly review the SPGPE formalism and introduce the full equation of motion. We also discuss the formal properties of the various parts of the SPGPE formalism individually, since it has been reasonably common to neglect various terms and noises in the equations of motion to arrive at simpler theories (e.g.~damped GPEs). While the growth process gives rise to a grand canonical description of the $\rC$-region, we show that scattering processes (without growth) realize a canonical description. By only including the scattering process damping term (i.e.~neglecting the associated noise) we arrive at an \textit{energetically damped} GPE  that evolves any initial field to the zero temperature ground state by removing energy, but not population. We also evaluate the coefficients of the scattering term and compare it to that of the growth term in the experimental regime, showing the coefficients to be comparable in size.

In Sec.~\ref{SecBreathingMode} we present our main tests of the formalism, and study the evolution of a condensate initial excited into a large amplitude breathing mode. We give a number of analytic results that we have used to calibrate our numerical algorithm, and provide some insight into the effects of scattering processes. Finally we present a study of the breathing mode in an experimentally realistic regime and demonstrate that the inclusion of scattering processes causes the system dynamics to change in a very significant manner.

After concluding and surveying the prospects for the full SPGPE we provide a detailed overview of our numerical algorithm  in the Appendix. 

\section{SPGPE formalism}\label{Sec:SPGPEformalism}
Here we briefly overview the formalism of the SPGPE. Detailed derivation of the equation of motion can be found in Refs.~\cite{SGPEI, SGPEII, SGPEIII}.  
\subsection{The stochastic projected Gross-Pitaevskii equation}
The SPGPE is a c-field method \cite{Blakie08a} where our system is decomposed in terms of eigenstates of the single-particle Hamiltonian, 
\eq{Hsp}
{H_{\rm sp} = -\frac{\hbar^2\nabla^2}{2m} +V(\x),}
where $
V(\x) =  \tfrac{1}{2}m(\omega_x^2x^2+\omega_y^2y^2+\omega_z^2z^2 ),$
is  the harmonic confinement potential.
  The eigenstates of (\ref{Hsp}),  satisfying $H_{\rm sp} \phi_n(\x) = \epsilon_n \phi_n(\x)$, form a convenient basis of states. Here the shorthand $n$ represents all quantum numbers required to specify a single-particle state. We introduce a single-particle energy cutoff $\ecut$ which separates the system into a low energy $\rC$-region consisting of single-particle modes with eigenvalues satisfying $\epsilon_n \leq \ecut$.  The energy cutoff is chosen so that all single-particle modes in the \rC-region have an appreciable occupation number, of order unity. In this case we can describe the bosonic field for the \rC-region with a classical field \cite{Blakie08a}, $\cf(\x,t)$, which we represent as a sum over single-particle states
\eq{cf}{\cf(\x,t) = \sum_{n\in\rC} c_n(t)\phi_n(\x),}
where the summation is restricted to all single particle modes in the $\rC$-region.  The remaining high energy incoherent ($\rI$) region contains thermally occupied modes, and acts as a thermal reservoir for the \rC-region.

The \rI-region is assumed to be in equilibrium with a well defined temperature $T$ and chemical potential $\mu$, and in a semiclassical local-density treatment is described by the single particle Wigner function
\eq{Fdef}{F(\x,\kk) = \frac{1}{\exp{[(\hbar \omega(\x,\kk) - \mu) /k_BT]-1 } },}
where the energy is $\hbar\omega(\x,\kk)=\hbar^2\kk^2/2m+V(\x)$.  

Accounting for the reservoir interactions with the \rC-region leads to the non-local Stratonovich stochastic equation of motion for $\cf(\x,t)$, known as the SPGPE (writing $\psi\equiv\psi(\x,t)$ for brevity) \cite{SGPEI,SGPEII,SGPEIII}
\eqn{\label{spgpe}
(S)\,\,d \cf&=&d \cf \big|_{H} +d \cf\big|_{G} +(S)\,\,d \cf \big|_{M},
}
where
\eqn{d \cf\big|_H &\equiv&{\cal P} \left\{ - \frac{i}{\hbar} {\cal L}\cf dt \right\},\label{hamil}\\
d \cf\big|_G &\equiv&{\cal P} \left\{ \frac{{G}(\x)}{k_B T} (\mu - {\cal L}) \cf dt + dW_G (\x,t)\right\},\label{growth}\\
(S)\,\,d \cf\big|_M&\equiv&{\cal P} \left\{ - \frac{i}{\hbar}V_M(\x,t)\cf dt +  i\cf dW_M(\x,t)\label{scatt}\right\},
}
with $(S)$ denoting Stratonovich integration~\cite{SM}. 
In Secs.~\ref{sssecHam}-\ref{sssecscat} we define the terms given in Eqs.~(\ref{hamil})-(\ref{scatt}), respectively.
A common feature to all terms is the projection operator  defined by
\eq{projector}
{\PC f(\x) \equiv \sum_{n \in \rC} \phi_n(\x) \intV{\x^{\prime}} \phi_n^*(\x^\prime)f(\x^\prime),}
which formally restricts the evolution of $\psi$ to the \rC-region.

\subsubsection{Hamiltonian evolution term: $d \cf \big|_{H}$}
The Hamiltonian evolution operator for the $\rC$-region,
${\cal L}$, is the usual GPE operator
\eq{LC}{
{\cal L}\cf \equiv (H_{\rm sp} + u|\cf|^2) \cf,
} 
where  $u =  4 \pi \hbar^2 a/m$ with $a$ the $s$-wave scattering length. However, as it appears projected in Eq.~(\ref{hamil})  this equation by itself is referred to as the projected GPE (PGPE). The PGPE contains the important interactions between the low energy modes as schematically indicated in   \fref{intFig}(a).

\subsubsection{Growth term: $d \cf\big|_{G}$}\label{sssecHam}
\begin{figure}[!t]
\begin{center}
\includegraphics[width=\columnwidth]{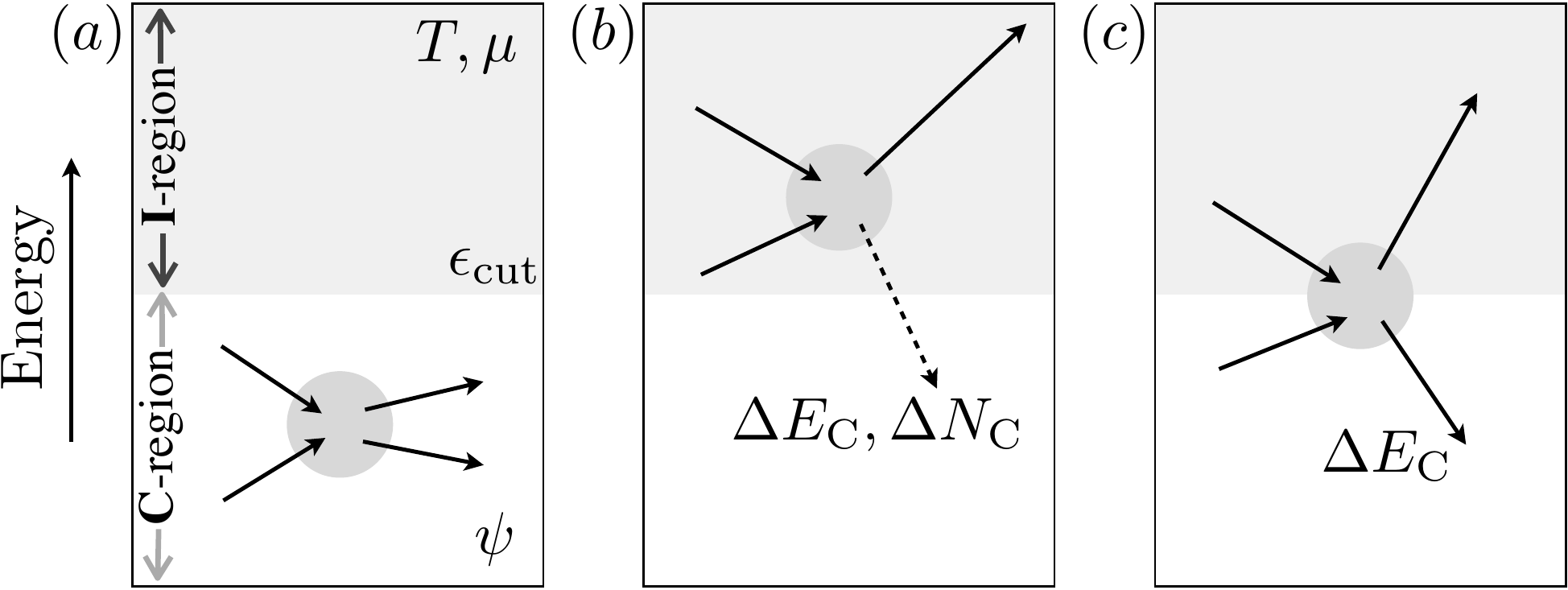}
\caption{The different interparticle scattering processes arising from each term of the SPGPE (\ref{spgpe}), where $\ecut$ specifies the boundary between \rC~and \rI.  The \rI-region is static, parameterized by the temperature ($T$) and chemical potential $\mu$.  (a) Non-linear mixing interaction between modes in the $\rC$-region. (b) Growth process in which both energy and number are transferred between \rC~and \rI. (c) Scattering process in which energy is transferred between \rC~ and \rI.  }
\label{intFig}
\end{center}
\end{figure}

The growth process [Eq. \eref{growth}]  describes the collisional interaction in which two $\rI$-region atoms collide leading to population growth of the $\rC$-region   
[see \fref{intFig}(b)]. It is set by the growth rate $G(\x)$ (see Sec.~\ref{ssecgrowth})  and the Gaussian complex noise $dW_G(\x,t)$ has the non-zero correlation
\eq{dWg}
{\langle dW_G^*(\x,t) dW_G(\x^\prime,t)  \rangle = 2G(\x) \delta_\rC (\x,\x^\prime) dt,}
where $\delta_\rC(\x,\x^\prime) = \sum_{n\in\rC}\phi_n(\x)\phi_n^*(\x^\prime)$ is a delta function in the $\rC$-region.   

\subsubsection{Scattering term: $d \cf\big|_{M}$}\label{sssecscat}
The scattering process [Eq.~\eref{scatt}] transfers energy and momentum between the $\rC$ and $\rI$-region without population transfer [see \fref{intFig}(c)]. It is set by the rate $M(\x)$ (see Sec.~\ref{ssecscattering}) which determines the effective potential 
\eq{vm}{
V_{M}(\x)=-\frac{\hbar^2}{k_B T}\mint{\x^\prime}M(\x-\x^\prime)\nabla^\prime\cdot\mathbf{j}(\x^\prime),
}
that couples the divergence of the $\rC$-field current 
\eq{jgp}{
\mathbf{j}(\x)=\frac{i\hbar}{2m}\left[\cf\nabla\cf^*-\cf^*\nabla\cf\right],
}
to the $\rI$-region.
The scattering noise $dW_{M}(\x,t)$ is real, defined by its non-vanishing correlation
\eq{dwm1}{
\langle dW_{ M}(\x,t)dW_{ M}(\x^\prime,t)\rangle=2M(\x-\x^\prime)dt.
}
The scattering noise forms a real-valued stochastic potential and thus generates a phase-diffusion process for the \rC-field evolution
that transfers energy and momentum between the $\rC$ and $\rI$-region atoms while preserving their populations.

\subsection{Formal properties of the full SPGPE } 

Here we briefly overview the formal properties of the full SPGPE theory.
Due to the coupling of the \rC-region with the reservoir, the SPGPE [Eq. (\ref{spgpe})] gives a grand canonical description of the \rC-region.  Irrespective of the form of the functions $G(\x)$ and $M(\x)$ (providing $G(\x) \neq 0$), the SPGPE evolves arbitrary initial conditions towards samples of the grand canonical ensemble, with equilibrium probability $P(\cf) \propto\exp(-K_\rC/k_BT)$, where $K_\rC \equiv E_\rC - \mu N_\rC$ is the grand canonical Hamiltonian, with
\begin{align} 
E_\rC&=\mint{\x}\cf^* H_{\rm sp}\cf +\frac{u}{2}\mint{\x}|\cf |^4,\label{Ec}\\
N_\rC&= \intV{\x} |\cf|^2,
\end{align}
the $\rC$-region energy and atom number, respectively.

In the next three subsections we examine the various sub-theories of the full SPGPE obtained by neglected various terms. This gives some insight into the role of the individual terms. Some of these sub-theories have been quite extensively used in the field,  particularly the established PGPE [Sec.~\ref{ssecPGPE}] and simple growth SPGPE [Sec.~\ref{ssecgrowth}]. The scattering SPGPE sub-theory presented in Sec.~\ref{ssecscattering} has not been considered before. In Table \ref{tab:methods} we present a summary overview of the various sub-theories considered and their general properties.

\begin{table*}[t!]
\begin{center}
\begin{tabular}{| c | c | c |  c | c|}
\hline\hline \multicolumn{2}{|c|}{\textbf{Method}} &\hspace{.0cm} \textbf{Ensemble} \hspace{.0cm} &\hspace{.0cm} \textbf{Conserved quantities} & \textbf{Variable quantities}\\
\hline\hline
\multicolumn{2}{|c|}{PGPE [Eq.~\eref{hamil}]} & Microcanonical & $E_\rC, N_\rC$ & -  \\
\hline 
 \multirow{2}{*} {Simple growth SPGPE [Eq. \eref{simpGrow}]} &$\emph{quiet}$ & NA & - &  $E_\rC, N_\rC$  \\ \cline{2-5}
 &\emph{noisy} & Grand-canonical & - &  $E_\rC, N_\rC$\\
\hline
 \multirow{2}{*} {\color{red}{\bf Scattering SPGPE  [Eq. \eref{scattspgpe}] }} &$\emph{quiet}$ & NA & $N_\rC$ &  $E_\rC$ \\ \cline{2-5}
 &\emph{noisy}& Canonical &  $N_\rC$ &  $E_\rC$\\
\hline
\multicolumn{2}{|c|}{\color{red}{\bf (full) SPGPE [Eq. \eref{spgpe}]}} & Grand-canonical & - &  $E_\rC, N_\rC$ \\
\hline\hline
\end{tabular}
\caption{Summary of the different theories considered in this paper. {\em Quiet} implies that the appropriate noise term is neglected, giving the damped PGPE ({\em quiet} simple growth SPGPE) and energetically damped PGPE ({\em quiet} scattering SPGPE).  The different reservoir processes leading to the distinct ensemble descriptions are summarized in \fref{intFig}. We emphasize that the scattering SPGPE enables a {\em dynamical} canonical description of the c-field.  The c-field methods which are implemented for the first time in this work are shown in bold red font.}
\label{tab:methods}
\end{center}
\end{table*}

\subsection{Projected GPE}\label{ssecPGPE}
The PGPE, defined by Eq.~\eref{hamil}, is formally (and numerically) number and energy conserving, for {\em any} finite cutoff $\ecut$ and single-particle basis. Since $\PC\PC=\PC$, we have $\PC\psi=\psi$ and 
\eqn{\label{pgpeN}
\frac{dN_\rC}{dt}\Big|_H&=&\int d^3\x\; \psi^*{\cal P} \left\{ - \frac{i}{\hbar} {\cal L}\cf dt \right\}+{\rm h. c.}\\
&=&\int d^3\x\; ({\cal P}\psi)^* \left\{ - \frac{i}{\hbar} {\cal L}\cf dt \right\}+{\rm h. c.}\label{pgpeNconj}\\
&=&0
}
where to get \eref{pgpeNconj} we have used the fact that $\PC$ is Hermitian~\cite{Bradley2005b}. Similar (lengthier) reasoning gives $dE_\rC/dt\big|_H=0$. While a formally Hamiltonian theory, the nonlinear interactions generate ergodic dynamics and the equation samples the microcanonical ensemble in equilibrium~\cite{Davis2003}. The PGPE describes the dynamics of both thermal and coherent $\rC$-region atoms non-perterbatively, giving a quantitatively accurate description of finite temperature systems in (or near) equilibrium where the reservoir interaction can be neglected \cite{Davis2006a}.  Hence the PGPE has mainly been applied to the equilibrium properties of the finite temperature Bose gas \cite{Davis2001b,Davis2002,Blakie05a,Davis2006a,Simula2006a,Bezett08a,Bezett09b,Bisset09d,Bisset09c,Wright09a,Wright10d,Wright11a}. Dynamical studies have considered vortex nucleation~\cite{Wright08a} and collective modes~\cite{Bezett09a}.

Numerically, the PGPE is a fully dealiased spectral method for propagating the GPE. In the basis of plane waves (i.e. the Fourier spectral method), the PGPE eliminates all spurious aliasing generated by four-wave interactions. In practice this is achieved by evaluating the interaction term for a wave function of $n$ points per spatial dimension ($n$ momentum modes) on a grid with $2n$ points in each dimension (extending out to twice the momentum cutoff of the wavefuction)~\cite{Davis2002}. This procedure is easily generalized to other bases~\cite{Blakie05a,Bradley08a}. 
\subsection{(Simple) Growth SPGPE}\label{ssecgrowth}
The only form of the SPGPE (\ref{spgpe}) which has been used for numerical simulations is the \emph{simple growth SPGPE}, in which scattering processes are neglected and the growth rate $G(\x)$ is taken as spatially uniform. The resulting equation is easily handled numerically and closely connected to the Ginzburg-Landau $\phi^4$ theory.

\subsubsection{Growth rate}
When the $\rI$-region is near equilibrium and described using Eq. \eref{Fdef}, $G(\x)$ is approximately spatially constant over the condensate \cite{Bradley08a}.  In this case the growth amplitude can be calculated explicitly as \cite{Bradley08a}
\begin{equation}\label{gamdef} G(\x)  \approx \gamma=\gamma_0\sum_{j=1}^\infty\; \frac{e^{\beta\mu(j+1)}}{e^{2\beta\ecut j}}\Phi\left[\frac{e^{\beta\mu}}{e^{\beta\ecut}},1,j\right]^2,
\end{equation}
where $\gamma_0 = 8a^2/\lambda_{dB}^2$, with $\lambda_{dB}=\sqrt{2\pi\hbar^2/mk_BT}$,  $\beta=1/k_BT$ and $\Phi[u,v,w]$  the Lerch transcendent. 
\subsubsection{Evolution equation}
The simple growth SPGPE is given by
\eqn{d \cf\big|_{H+\gamma}&=&d\psi\big|_H+d\psi\big|_\gamma,\label{simpGrow}
}
where
\eqn{
d\psi\big|_\gamma&\equiv&\PC\left\{\frac{\gamma}{k_B T} (\mu - {\cal L}) \cf dt + dW_\gamma (\x,t)\right\} \label{simpleSPGPE} ,}
and the noise correlation is
\eq{dWsimpleg}
{\langle dW_\gamma^*(\x,t) dW_\gamma(\x^\prime,t)  \rangle = 2\gamma \delta_\rC (\x,\x^\prime) dt.}

The  numerical integration of (\ref{simpleSPGPE}) is of a similar computational expense to solving the PGPE \cite{Blakie08b} since the noise is additive and weak, and we can use a higher order Runge-Kutta algorithm to achieve stochastic convergence \cite{Milstein}. 

\subsubsection{Properties of the simple growth SPGPE}
Without noise ($dW_\gamma\equiv 0$, a case we call {\em quiet}), the simple growth SPGPE reduces to the \emph{damped PGPE}  \footnote{The damped PGPE is formally similar to the damped GPE \cite{Choi1998} which was applied to phenomenological studies of vortex lattice formation \cite{Penckwitt2002,Tsubota2002,Kasamatsu2003}.}. 
The damped PGPE evolves $N_\rC$ to equilibrium:
\eqn{\label{dNdtsge}
\frac{dN_\rC}{dt}\Big|_{H+\gamma,quiet}=-2\gamma[\tilde{\mu}(t)-\mu]N_\rC,
}
where $\tilde{\mu}(t)=\int d^3\x\psi^*{\cal L}\psi/N_\rC$ is the instantaneous chemical potential.
This evolution also causes energy to decay uniformly:
\eqn{\label{dKdtsge}
\frac{dK_\rC}{dt}\Big|_{H+\gamma,quiet}=-\frac{2\hbar\gamma}{k_BT}\int d^3\x |(\mu-{\cal L})\psi|^2,
} 
thus minimizing $K_\rC$ and damping out thermal fluctuations.  The equilibrium solution is the zero temperature ground state of the PGPE \eref{hamil} satisfying $\mu\psi_0=\PC\{{\cal L}\psi_0\}$. 

Upon reintroducing the noise, Eq.~\eref{simpGrow} samples the grand canonical ensemble and $K_\rC[\psi]>K_\rC[\psi_0]$ for any sample $\psi$. However, all equilibrium properties are independent of the choice of $\gamma$. Confirming this property provides an excellent test of any numerical implementation.

\subsection{Scattering SPGPE}\label{ssecscattering}

We define the \textit{scattering} SPGPE to be the sub-theory of the SPGPE obtained by neglecting growth terms.

\subsubsection{Scattering rate}
The scattering rate is given by
\eq{Mdef}{
M(\x)=\frac{{\cal M}}{(2\pi)^3}\mint{\kk}\frac{e^{i\kk\cdot\x}}{|\kk|},
}
where
\eq{Mfac}{
{\cal M}\equiv\frac{16\pi a^2k_BT}{\hbar}\frac{1}{e^{\beta(\ec-\mu)}-1}.
}
In Ref.~\cite{SGPEII} this rate was referred to as the ``simplified non-local form", however it can be shown (see Appendix \ref{secscaterateapp}) that, within the semiclassical approximation used in the Wigner function $F(\x,\mathbf{k})$ description of the $\rI$-region, it is exactly equal to the full non-local form.

\subsubsection{Evolution equation}
The {scattering SPGPE}  takes the form
\eqn{
(S)\,\,d\cf\big|_{H+M} &=&d\psi\big|_H+(S)\,\,d\psi\big|_M.\label{scattspgpe}} 
As the scattering process is fundamentally number-conserving, the scattering SPGPE provides a {\em dynamical} canonical description of the $\rC$-region.  Equilibrium states of \eref{scattspgpe} sample the canonical distribution with probability $P(\cf) \propto\exp(-E_\rC/k_BT)$. 

Numerically integrating the scattering SPGPE (\ref{scattspgpe}) poses a heavier technical challenge compared to evaluating the simple growth SPGPE (\ref{simpGrow}).  Due to the multiplicative noise arising from the scattering process, we are restricted to using a semi-implicit Euler algorithm to evolve the stochastic differential equation (see appendix section \ref{sec:Eulerevo}).  This algorithm is first order in the weak sense of convergence of Ref.~\cite{SM}, hence is much more inefficient (in computation resources) than the higher order Runge-Kutta algorithms that can be used to evolve the simple growth SPGPE. Further complexity arises through the non-local deterministic term, $V_{M}(\x)$ [Eq.~(\ref{vm})], and in sampling the correlated noise $dW_M$. 
 The key steps in our numerical implementation use techniques developed for solving the PGPE with dipole-dipole interactions  \cite{Blakie2009a}, and a full overview of the algorithm is given in Appendix \ref{appa}.

\subsubsection{Properties of the scattering SPGPE}
Setting $dW_M \equiv 0$ in Eq.~(\ref{scattspgpe}) leads to the \emph{quiet} form of the scattering SPGPE, which we call the \emph{energetically damped PGPE}. The deterministic part of the scattering term $V_M$ takes the form of an effective potential  that acts to reduce the energy of the $\rC$-region, Eq.~\eref{Ec}.  Using Eq.~(\ref{scattspgpe}) with $dW_M\equiv 0$ we find
\eq{energy}{
\frac{dK_\rC}{dt}\bigg|_{H+M, quiet}=\frac{dE_\rC}{dt}\bigg|_{H+M, quiet}=\mint{\x}V_M(\x)\nabla\cdot\mathbf{j}(\x).
}
Substituting Eq.~(\ref{vm}) and Eq.~(\ref{Mdef}) gives
\eqn{
\frac{dK_\rC}{dt}\bigg|_{H+M, quiet}&=&-\frac{\hbar^2{\cal M}}{k_BT}\int \frac{d^3\kk}{|\kk|}\left|\kk\cdot\mathbf{j}(\kk)\right|^2   \label{dHdtpos}}
which is negative semi-definite. The scattering term causes the $\rC$-region energy to monotonically decrease, which provides a useful consistency check on the the accuracy of the numerical evaluation of $V_M$.  

The Ehrenfest relation (\ref{energy}) gives an important physical insight into the role of the scattering process in the SPGPE. The scattering process generates a dissipative interaction that enters the evolution as an effective {\em Hamiltonian} term, in the form of a stochastic effective potential. When the noise is neglected, energy is continually removed through evolution [Eq.~(\ref{dHdtpos})] and the system proceeds toward the absolute ground state. The inclusion of the noise ensures that the effective potential is stochastic, maintaining the finite temperature character of the $\rC$-region. 

We have identified some basic tests that proved useful in validating our numerics.  First, equilibrium ensemble properties will be independent of the scattering coefficient ${\cal M}$ for a given temperature (much as the simple growth SPGPE equilibria are independent of $\gamma$), and should be equivalent to those generated by the simple growth SPGPE for the same final particle number (assuming equivalence of the canonical and grand canonical ensembles).  Second, we expect that a stochastic scattering term implementation will evolve $N_\rC$ particles in the \rC-region into thermal equilibrium. Evolution according to the deterministic term will evolve the c-field region toward the $N_\rC$-particle PGPE ground state. 

\begin{figure}[!t]
\begin{center}
\includegraphics[width=1\columnwidth]{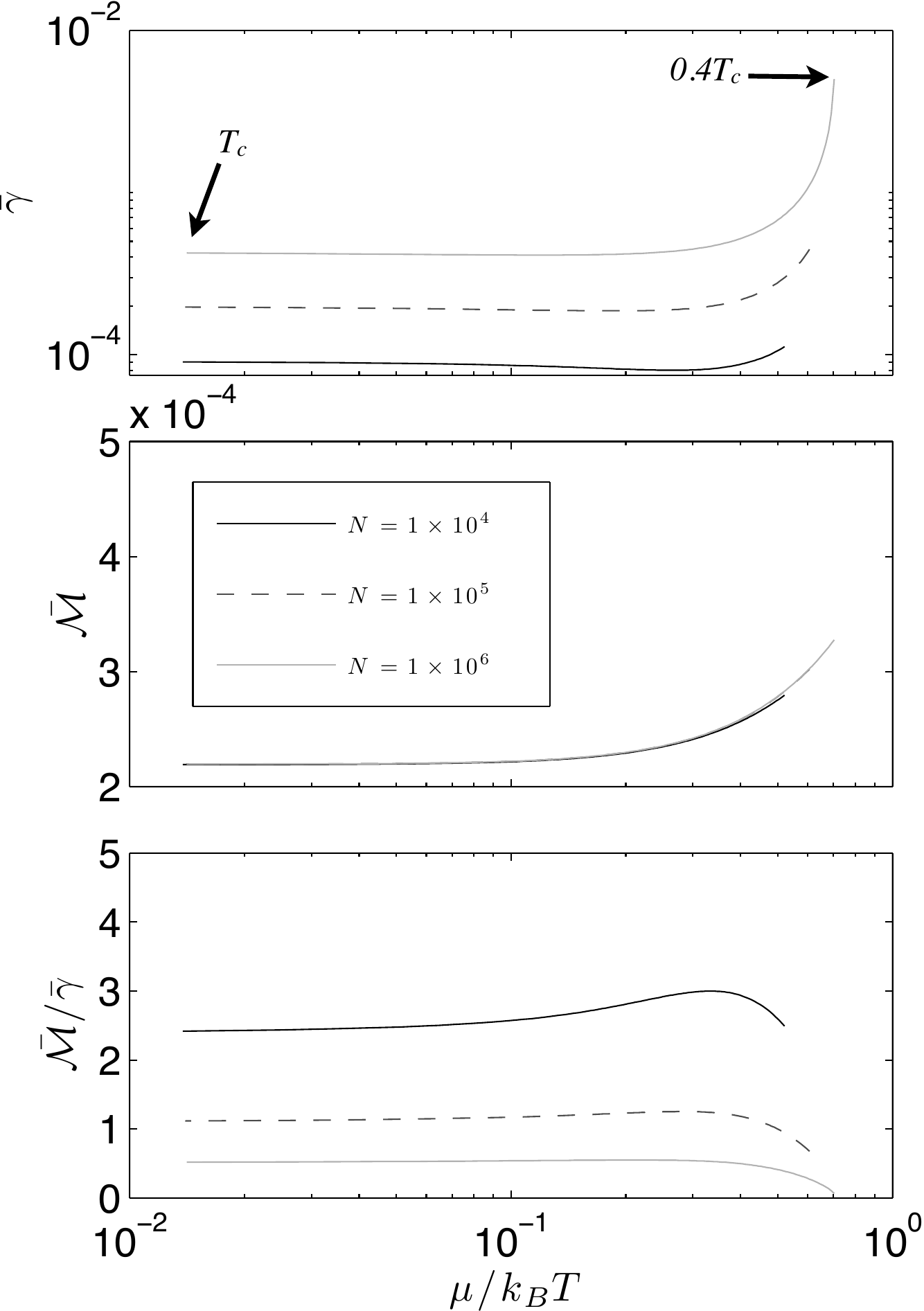}
\caption{The growth ($\bar\gamma$) and scattering ($\bar{\mathcal{M}}$) coefficients, and their ratio, as a function of $\mu/k_BT$.  Reservoir parameters $T,\mu,\ecut$ are determined from the Hatree-Fock parameter estimation scheme from Ref.~\cite{Rooney10a}.  The parameters found ensure $T$ can be varied without altering the total atom number $N$, and critical temperature $T_c(N)$.    Along each curve the temperature varies from $T = T_c$ to $T = 0.4T_c$, where $T_c$ corresponds to a total atom number of $N = 1\times10^4$ (black), $N = 1\times10^5$ (dashed gray), $N = 1\times10^6$ (gray).}
\label{fig:damp_paramcomp}
\end{center}
\end{figure}

\subsection{Comparison of the scattering and growth coefficients}
To assess the relative importance of scattering and growth terms in the SPGPE it is of interest to evaluate the respective coefficients ($ \gamma_0$ and $\mathcal{M}$) in regimes relevant to experiments. It is useful to consider these in their dimensionless forms
\begin{align}
\bar{\gamma}_0&\equiv \frac{\gamma_0\hbar}{k_BT},\\
\bar{\mathcal{M}} &\equiv \frac{\mathcal{M}\hbar}{k_BTx_0^2}\approx\frac{8\pi a^2}{x_0^2}\frac{k_B T}{\mu},\label{Mdl}
\end{align}
where $x_0 = \sqrt{\hbar/m\bar{\omega}}$ with $\bar{\omega} =\sqrt[3]{\omega_x\omega_y\omega_z}$. The approximation in Eq.~(\ref{Mdl}) is obtained
for the usual validity regime of the SPGPE theory \cite{Blakie08a} $k_BT\gg\mu$, and typical cutoff choice $\ec\sim 3\mu$. 
Thus the ratio of the coefficients is given by
\eq{Mapprox}{
\frac{\bar{\cal M}}{{\bar{\gamma}_0}}\approx \frac{\lambda_{dB}^2}{\pi x_0^2}\frac{k_B T}{\mu}.
}
In usual experimental regimes  $\lambda_{dB}\sim x_0$ (for temperatures near the critical temperature)  and thus we conclude that the scattering coefficient $\bar{\cal M}$ is significant, potentially appreciably exceeding the growth coefficient. 

To be more quantitative we evaluate the coefficients using a numerical calculation (see Ref.~\cite{Rooney10a}) for a spherically trapped system with $\omega_r = 2\pi\times10$~Hz and a range of total atom numbers $N = N_\rC + N_\rI$ (where $N_\rI$ is the number of atoms in the $\rI$-region) and temperatures ($\mu$ is found to ensure $N$ is fixed as $T$ varies).  The results for $\bar{\gamma}$ and $\bar{\mathcal{M}}$ are shown in \fref{fig:damp_paramcomp}, noting that for the growth term we evaluate the full coefficient $\gamma$ (\ref{gamdef}). These results support the qualitative analysis given above, and show the coefficients are similar in size over a broad regime.  However the {\em net} effect of scattering depends on the divergence of currents in the $\rC$-region, and can be small for quasi-equilibrium dynamics where the simple growth SPGPE has been successful \cite{Rooney2012a}.

\section{Breathing mode decay}\label{SecBreathingMode}
As an application of our implementation of the SPGPE we study the case of a BEC excited into a large amplitude breathing oscillation.

\subsection{Gaussian wave function}\label{sec:Gaussian}
For the results we consider in this section we begin with the well-defined initial condition for the $\rC$-region field:
\eq{Gaussianwf}{
\cf({\rr})=\frac{\sqrt{N_{\rC,i}}}{ (\pi\sigma^2)^{\frac{3}{4}}} e^{-r^2/2\sigma^2+i\kappa r^2/2},
} 
where $N_{\rC,i}$ is the initial $\rC$-region number, and $\sigma$ and $\kappa$ are real constants.   The $\rC$-region energy per particle of this initial field, found analytically by evaluating (\ref{Ec}), is
\eq{Eanalytic}{\frac{E_{\rC}}{N_{\rC,i}} = \frac{3\hbar^2}{4m}\left(\frac{1}{\sigma^2} + \sigma^2\kappa^2  \right) + \frac{3m\omega_r^2\sigma^2}{4} + \frac{uN_{\rC,i}^2}{\sqrt{32}\pi^{3/2}\sigma^3}.}

We begin by looking at a spherically trapped condensate, with a trapping frequency of $\omega_r = 2\pi\times10$~Hz, and choose parameters of the initial field to be $\sigma = x_0, \kappa = x_0^{-2}$, with $N_{\rC,i}  = 1\times 10^4$  $^{87}{\rm Rb}$ atoms.  For these parameters, we find from (\ref{Eanalytic}) that $E_0 \equiv E_\rC(t\!=\!0) =  8.44\hbar\bar{\omega}N_{\rC,i}$.  The ground state of the Gross-Pitaevskii equation for this system is $E_G  = 3.53 \hbar\bar\omega N_{\rC,i}$, so our initial state is far from equilibrium.  The current density of this initial condition is given by
\eq{jGauss}{
\mathbf{j}(\rr)=\rr\kappa|\cf|^2,}
i.e.~a radially expanding motion.  
Using   (\ref{jGauss}) we can evaluate \eref{vm} to find an analytic expression for $V_M$
\eq{Vmtot}{
{V}^A_M(r)=-\frac{\mathcal{N}^2\bar{\cal M}\kappa\sigma}{\sqrt{2}}g\left(r/\sigma\right) \hbar\bar{\omega},
}
where
\eq{gdef}{
g(x)=\sqrt{\frac{2}{\pi}}+\left(\frac{1}{\sqrt{2}x}-\sqrt{2}x\right)e^{-x^2}{\rm erfi}(x),
}
with ${\rm erfi}(x)\equiv -i{\rm erf}(ix)$. For $x\ll1$,  $g(x)$ has the asymptotic expansion
\eq{gsmallx}{
g(x)=\sqrt{\frac{8}{\pi}}\left(1-\frac{4}{3}x^2\right)+O(x^4),
}
while $\lim_{x\to\infty}g(x)=0$.  For our choice of initial condition ($\kappa>0$, i.e.~radial flow away from the origin) we have ${V}_{M}(0)<0$ and ${V}_{M}(r)\to0$ as $r\to\infty$. Thus for small $r$ $V_M(r)$ is approximately described as an additional harmonic potential well which acts against the radial expansion. In addition to  providing extra confinement, $V_M$ has a  negative energy offset, which modifies the effective energy minimum of the potential experienced by the system, and thus could effect growth into the condensate. 
 
\begin{figure}[!t]
\begin{center}
\includegraphics[width=0.85\columnwidth]{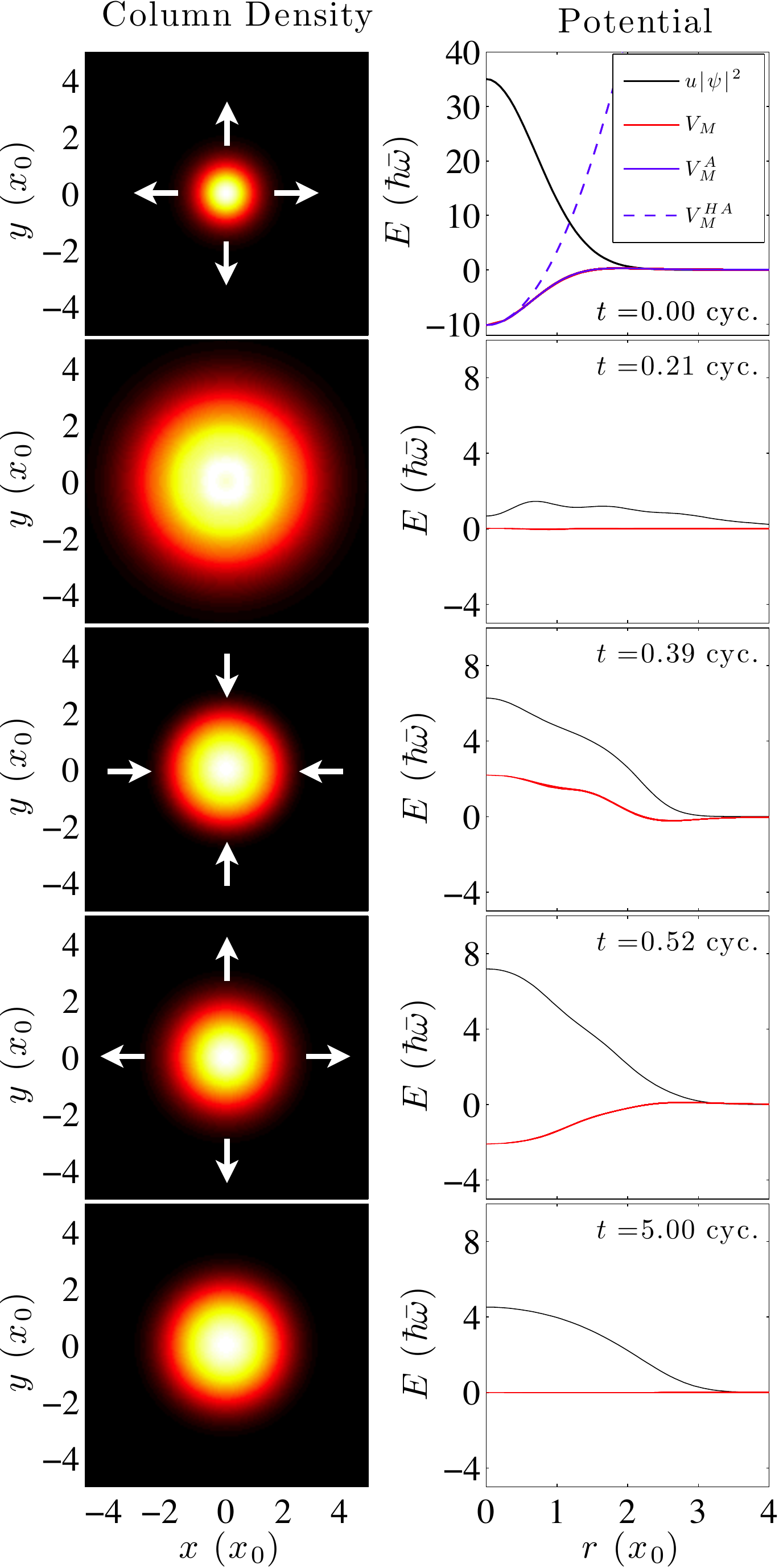}
\caption{Results of the energetically damped PGPE, comparing the density with the scattering effective potential where $\bar{\mathcal{M}} = 0.005$, at representative times (in units of trap cycles).  Column densities are shown in the left column, with arrows indicating the direction of the breathing motion and flow of current at each time.  The scattering potential, $V_M$ (red curve), and a radial slice of density in energy units, the local $\rC$-region interaction energy $u|\cf|^2$ (black curve), are shown in the right column.  At $t=0$ we show $V_M^A$ (blue curve), the analytical scattering potential for a Gaussian wavefunction found from (\ref{Vmtot}).  $V_M^{HA}$ (blue dashed curve), is (\ref{Vmtot}) found using the harmonic approximation in Eq.~(\ref{gsmallx}).}
\label{fig:seqGauss1e4}
\end{center}
\end{figure}

In Fig.~\ref{fig:seqGauss1e4} (upper right subfigure) we compare ${V}^A_M(r)$ with ${V}_M(r)$ obtained numerically using the initial condition \eref{Gaussianwf} and find excellent agreement.

\begin{figure}[!t]
\begin{center}
\includegraphics[width=\columnwidth]{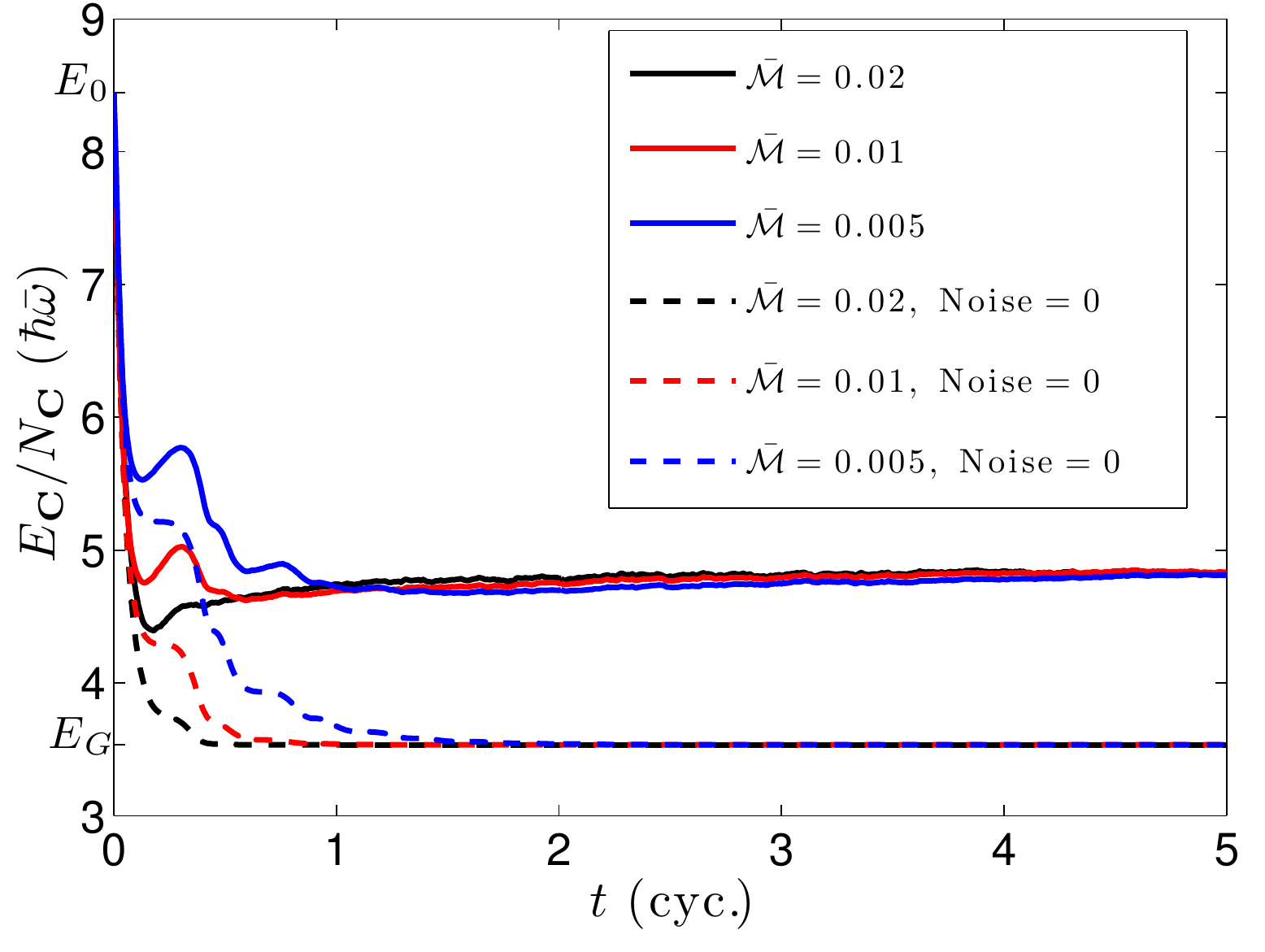}
\caption{\rC-field energy per particle as a function of time (in trap cycles), for the scattering SPGPE evolution of an initial Gaussian wave function with $N_\rC = 1\times10^4$ atoms.  (solid lines) scattering SPGPE evolution including the noise, at temperature of $T = 10 \hbar \bar\omega /k_B$ for a range of $\bar{\mathcal{M}}$.  (dashed lines) Evolution with the deterministic scattering term, neglecting the noise, for a range of $\bar{\mathcal{M}}$.  The initial energy ($E_0$) determined analytically from \eref{Eanalytic} agrees with the numerical calculation.  For the {\em quiet} simulations, the ground state energy agrees with the Gross-Pitaevskii equation ground state energy labelled $E_G$. For the {\em noisy} simulations, the occupation of thermally excited \rC-region modes raises the equilibrium energy.}
\label{fig:energy}
\end{center}
\end{figure}
\subsection{Scattering: deterministic versus noisy dynamics}
We examine the scattering SPGPE evolution of a Gaussian wave function (\ref{Gaussianwf}), with the parameters specified in section \ref{sec:Gaussian}.  First, we consider the effect of the scattering effective potential term on its own by using the energetically damped PGPE [Eq.~\eref{scattspgpe} without noise].  In \fref{fig:seqGauss1e4} we show the deterministic evolution of the Gaussian wave function, comparing the density with the effective potential at a range of representative times.  The arrows on the density images show the direction of the breathing motion (i.e.~the direction of current flow) at each time.  As the breathing motion evolves, $V_M(r)$  acts against the density change.  As the condensate expands we have $ V_M(r\approx0) < 0$, where the negative potential acts against radial expansion.  Then as the condensate begins to contract and the flow is directed towards the origin, we see $  V_M(r\approx0) > 0$ (see t = 0.39 cyc.).  This is consistent with Eq.~\eref{Vmtot} evaluated with $\kappa<0$ (i.e.~consistent with an inward current).  Finally the currents are completely damped out and the system reaches equilibrium, i.e.~the ground state of the PGPE (shown at $t = 5$ trap cycles).  

The evolution of the $\rC$-region energy for the energetically damped PGPE is shown in Fig.~\ref{fig:energy}. We see the expected behavior, namely that $V_M$ acts to reduce the energy monotonically until the system reaches the ground state, with energy consistent with the zero temperature GPE ground state ($E_G$).  The final state is independent of the value of $\mathcal{M}$, although this does influence the rate at which the final states are reached. We can quantify the initial effect that $V_M(r)$ has on the rate of change of the $\rC$-region energy. To do this we evaluate Eq.~(\ref{dHdtpos}) for the initial field (\ref{Gaussianwf}), giving 
\eq{dHdtanalytic}
{\frac{dE_\rC}{dt} = -\frac{\hbar^2 \mathcal{M}}{k_BT} \left(\frac{\hbar \kappa N_{\rC,i}}{\pi m \sigma} \right)^2, }
which describes the rapid loss of energy initially seen in  Fig.~\ref{fig:energy}. Using the parameters of our initial state, we find $dE_\rC/dt=-5.1\times 10^4\hbar\bar{\omega}^2$, which agrees with our numerical evaluation of Eq.~\eref{dHdtpos}  to better than one part in $10^4$.

We now examine the role of the scattering noise term, for the same parameters as above, and a temperature of $T = 10 \hbar\bar\omega/k_B$.  The energetic evolution for this case is also shown in Fig.~\ref{fig:energy}.  For the stochastic simulations, we have averaged over 10 trajectories for each parameter set.  From the initial condition we again observe energy to decay, however unlike the noiseless simulations where it strictly decreases, we see that this is not the case  for the noisy simulation. The local peaks in $E_{\rC}$ occur when the condensate is fully contracted in its breathing motion. We observe that, as expected, the final finite temperature equilibrium state is independent of $\mathcal{M}$. The equilibrium states has appreciably more energy than the ground state, reflecting the thermal excitation of the system.

\subsection{Finite temperature breathing mode decay in an experimentally realistic regime}
Finally we extend our study of the breathing mode to a regime with experimentally realizable parameters and compare the predictions of the full, scattering, and simple growth SPGPEs.

\subsubsection{Parameter choice}
 In order to give a well defined comparison, we choose physically consistent reservoir parameters using the Hartree-Fock parameter estimation scheme described in Ref.~\cite{Rooney10a}.  For a  total atom number of $N = N_{\rC,i}+N_\rI = 5\times10^4$ atoms, we find $T = 29.4 \hbar\bar\omega/k_B$, $\mu = 4.8\hbar\bar\omega$, $\ec= 15.9\hbar\bar\omega$ ($T \approx 0.85 T_c$), giving   $(\bar\gamma,\bar{\mathcal{M}}) = (1.5,2.7)\times10^{-4}$ [from Eqs.~(\ref{gamdef}) and (\ref{Mdl})].   
Our initial state consists of a Gaussian field with radial phase gradient (\ref{Gaussianwf}), with the same parameters as specified in section \ref{sec:Gaussian},  except with $N_{\rC,i} = 1.22 \times 10^4$, which corresponds to the Thomas-Fermi condensate number from our value of  $\mu$. We evolve this initial state with the same reservoir parameters $T$, $\mu$, and $N_\rI$ determined above.

\begin{figure}[!t]
\begin{center}
\includegraphics[width=\columnwidth]{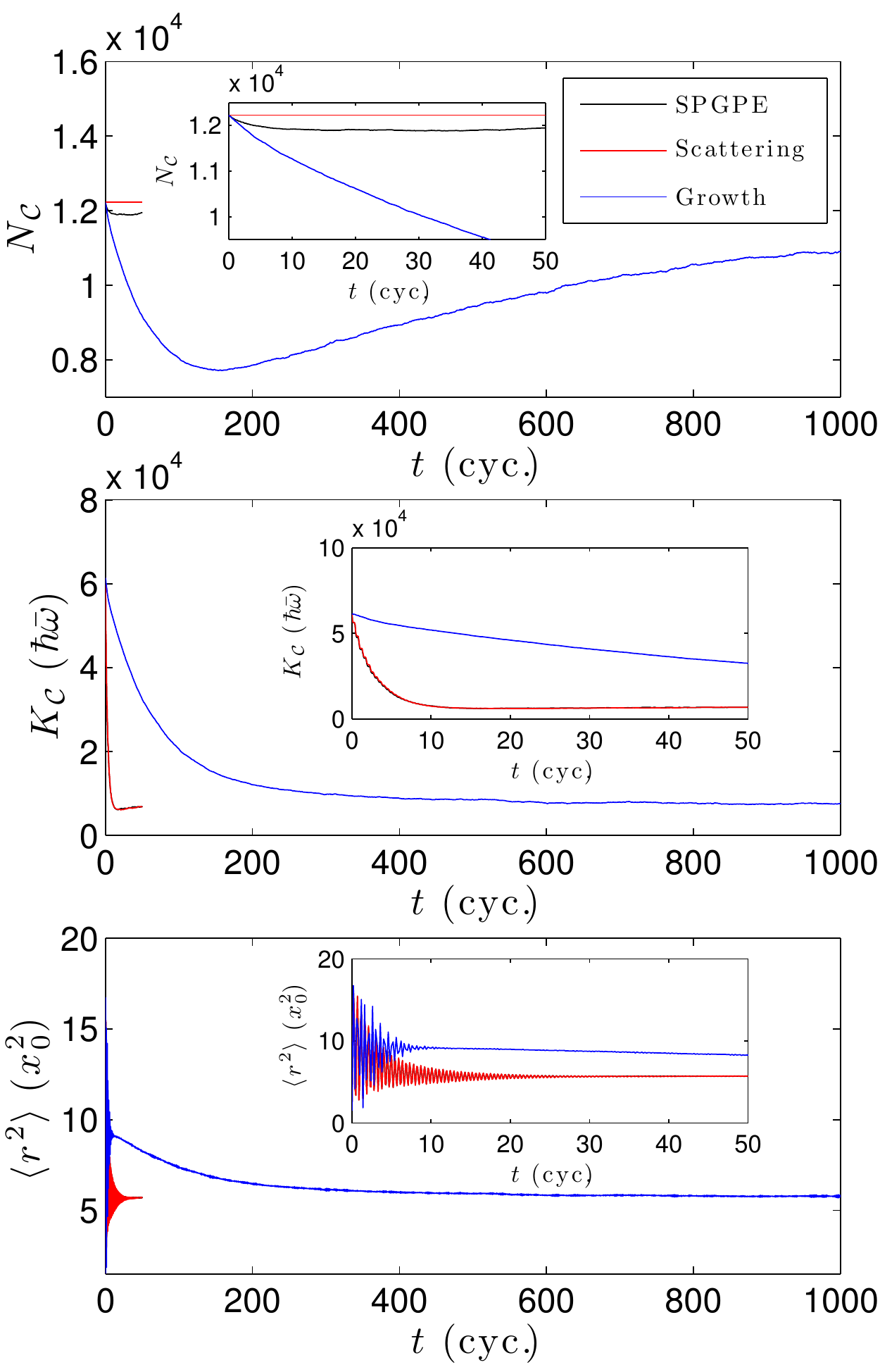}
\caption{$N_\rC$, $K_\rC$, and $\langle r^2 \rangle$ as a function of time, for the evolution of a Gaussian wave function using the SPGPE (black), scattering SPGPE (red), and simple growth SPGPE (blue).  We show the time evolution for each method until equilibrium has been reached.  In the inset we focus on the early time dynamics, over which the SPGPE and scattering SPGPE reach equilibrium at a much faster rate than the simple growth SPGPE.  The SPGPE and scattering SPGPE results for $K_\rC$ and $\langle r^2 \rangle$ are virtually indistinguishable.}
\label{fig:comparison}
\end{center}
\end{figure}

\subsubsection{Comparison of evolution}
In \fref{fig:comparison} we compare the evolution of $N_\rC$ and $K_\rC$ as an average of 50 trajectories, throughout the decay of the breathing motion.  The number and energy reach equilibrium much faster for the full and scattering SPGPEs than for the simple growth SPGPE.  The behavior of $K_{\rC}$ is almost identical for the full and scattering SPGPEs, with the difference in the value of $N_\rC$ that these two theories equilibrate to arising because the scattering SPGPE conserves number.  
In contrast, the simple growth SPGPE predicts very different behavior:  $K_\rC$ changes in a similar way to the other theories, but on a much slower timescale. $N_\rC$ instead evolves in a different manner, decreasing to about 80\% of $N_{\rC,i}$ before slowly returning towards this initial value (an equilibrium value of $N_{\rC}=1.15\times 10^{4}$ is eventually reached).
Note we have verified that our numerical algorithm produces the same equilibrium state (after sufficiently long times) for the full and simple growth SPGPEs, irrespective of the value of $\bar\gamma$ and $\bar{\mathcal{M}}$.  

To quantify the decay of the actual breathing mode we calculate $\langle r^2 \rangle$, which provides a measure of the average system width (noting $\langle r \rangle \approx 0$).  The evolution of $\langle r^2 \rangle$ is shown in \fref{fig:comparison}, with  the initial oscillations of $\langle r^2 \rangle$ correspond to the breathing mode of the condensate. We see the damping of $\langle r^2 \rangle$  for all methods is broadly consistent with that of $K_\rC$.  However $\langle r^2 \rangle$ for the simple growth SPGPE simulations shows an interesting difference:  The oscillations decay  within 10 trap cycles, despite $\langle r^2 \rangle$ (as well as $N_\rC$ and $K_\rC$) being far from equilibrium. After this time $\langle r^2 \rangle$ decays slowly (without oscillation) towards the equilibrium value.

\subsubsection{Condensate dynamics}
\begin{figure}[!t]
\begin{center}
\includegraphics[width=\columnwidth]{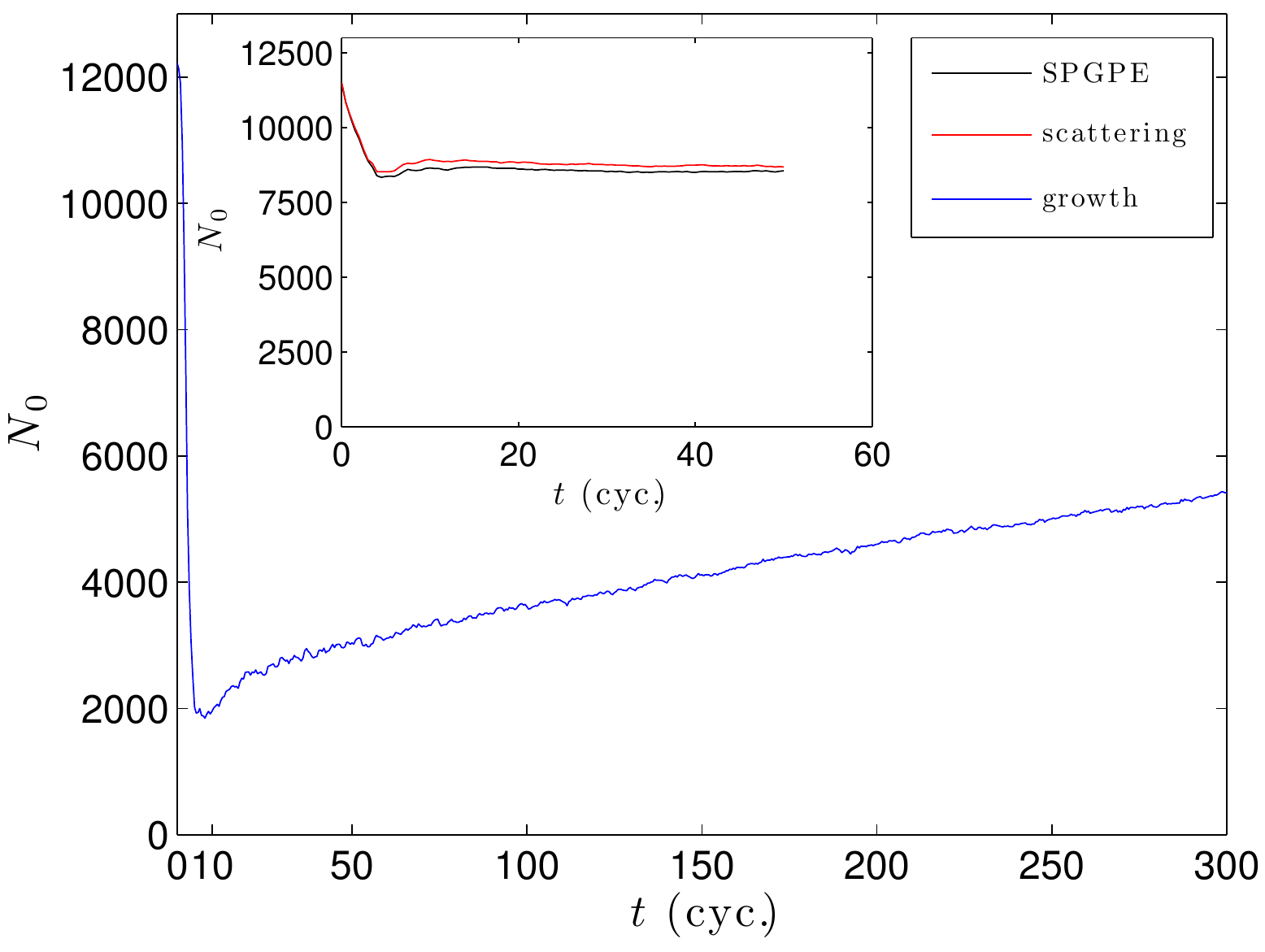}
\caption{Condensate fraction found from the Penrose-Onsager criterion, as a function of time for the same system as in \fref{fig:comparison}. The SPGPE (black) and scattering SPGPE (red) simulations are shown in the inset, and the simple growth (blue) simulations are shown in the main figure.}
\label{fig:condensate}
\end{center}
\end{figure}

To understand the marked difference between the simple growth evolution and the other two theories it is useful to consider the behavior of the condensate, which we examine in \fref{fig:condensate}.
We determine the condensate number from $\rC$-region field using the Penrose-Onsager definition \cite{Penrose1956}:  We form the one-body density matrix 
\eq{denmat}{
\rho_1(\x,\x^\prime,t) =\langle \cf^*(\x,t) \cf (\x^\prime,t) \rangle,
}
where angle-brackets denote an ensemble average over trajectories at a given time $t$. The condensate number $N_0(t)$ is defined as the largest eigenvalue of $\rho_1(\x,\x^\prime,t)$, and in our calculations always greatly exceeds the next largest eigenvalue.
 Our initial state for the $\rC$-region is essentially a pure condensate, with $N_0=N_\rC$ at $t=0$.  Similar to the observations made of \fref{fig:comparison}  we see that $N_0$ reaches equilibrium far more rapidly for the full and scattering SPGPEs compared with the simple growth SPGPE. Our results show that in the simple growth SPGPE the condensate fraction rapidly drops (over a time period consistent with the rapid decay in $\langle r^2\rangle$) to a minimum condensate number of $N_0 = 1.85\times10^3$ at $t = 8$ cycles.  The very slow approach to equilibrium observed after this time corresponds to a re-condensation process, as can be seen in the long-time evolution of $\langle r^2\rangle$ in Fig.~\ref{fig:comparison}(c). 
 The scattering term (i.e.~full and scattering SPGPEs) allows a different and very effective route to rapidly dissipate the energy of the breathing mode without drastically reducing the condensate fraction.

%
\section{Conclusions} 

When the full SPGPE theory was derived by Gardiner and Davis in 2003 they touted that ``This approach is distinguished by the control of the approximations made in its derivation, and by the feasibility of its numerical implementation" \cite{SGPEII}. In this work, we have finally realized a numerical implementation of this theory, and demonstrated practical simulations in the experimental regime.  

  To date all applications of the SPGPE have been made within the simple growth approximation in which the scattering terms are neglected. Using our algorithm we are able to assess the effects of the scattering terms.  We have verified that, when growth terms are neglected, the scattering terms evolve the system to an equilibrium state that is independent of the scattering amplitude coefficient (${\cal M}$), and that samples the canonical ensemble for the $\rC$-field region. The latter property is distinct to the simple growth and full SPGPE descriptions that exchange both energy and particles with the reservoir, and hence sample the grand canonical ensemble in equilibrium. 
 
 We have applied our theory to study the evolution of a finite temperature condensate excited into a large amplitude breathing mode in a physically realizable regime. An important, and somewhat surprising observation, is that the SPGPE with the scattering terms predicts a  qualitatively different evolution to the simple growth SPGPE: with the inclusion of  scattering, the breathing oscillation is efficiently damped without greatly depleting the condensate, allowing equilibrium to be established on a much shorter time scale. The energy damping is  due to the scattering effective potential that precisely opposes superfluid motion, and the results suggest that the scattering describes coherent energy exchange with the reservoir, a striking consequence of Bose-enhancement.  Our results indicate that scattering terms may be important in highly non-equilibrium regimes encountered in ultra-cold gases, and that quantitative evidence for the dominance of scattering over growth might be easily measured in experiments.

Future work with the full SPGPE will be to advance our understanding of non-equilibrium dynamics in the finite temperature regime. An exciting prospect is the direct comparison with experiments of a non-equilibrium scenario of finite temperature dynamics, such as the breathing mode decay studied here, or the dynamics of condensate growth during a quench~\cite{Weiler08a,Smith2012a}.

\section*{Acknowledgments}
We thank C.~W.~Gardiner and U.~Z\"ulicke for valuable discussions.  We acknowledge the use of the University of Otago Vulcan cluster, and the Victoria University of Wellington Sci-Fac HPC facility.  This research was supported by the Marsden Fund of New Zealand contracts UOO0924 (PBB), and UOO162 (ASB), and the Royal Society of New Zealand (contract UOO004). SJR was supported by the University of Otago, and thanks the Victoria University of Wellington for their hospitality. 
\appendix
\section{Scattering Rate}\label{secscaterateapp}

The full non-local scattering rate, as derived in \cite{SGPEII} is of the form 
\eq{Mgard}{
M(\mathbf{R},\x)=\frac{16\pi a^2k_BT}{(2\pi)^3\hbar}\mint{\kk}\frac{e^{i\kk\cdot\mathbf{r}}}{|\kk|}\frac{1}{e^{\beta(E_{\rm min}(\mathbf{R})-\mu)}-1},
}
 where
 \begin{equation}
 E_{\rm min}(\mathbf{R})\equiv{\rm max}\left(\ec\,,\,\frac{\hbar^2|\mathbf{k}|^2}{8m}+V_{\rm eff}(\mathbf{R})\right).
 \end{equation}
This result is obtained as the Fourier transform of 
 \eqn{
\tilde{M}(\mathbf{R},\kk)&=&\frac{4\pi u^2}{\hbar^2}\int_{\rI} \frac{d^3\mathbf{K}_1}{(2\pi)^3}\int_{\rI} \frac{d^3\mathbf{K}_2}{(2\pi)^3}\delta(\mathbf{K}_1-\mathbf{K}_2-\kk) \nonumber \\
&\times&\delta(\omega_1-\omega_2)F(\mathbf{R},\mathbf{K}_1)[1+F(\mathbf{R},\mathbf{K}_2)],\label{Mfull}
}
where $\mathbf{K}_1$ and $\mathbf{K}_2$ are the wavevectors of the $\rI$-region atom before and after a scattering collision, respectively [see Fig.~\ref{intFig}(c)]. The $\rI$ subscripts on the integrals in Eq.~(\ref{Mfull}) indicate the domain of integration is restricted to the $\rI$-region, where $\hbar\omega(\mathbf{R},\kk)>\ec$.

 This rate can be further simplified as follows: In a semiclassical description conservation of momentum during the scattering collision requires $\mathbf{K}_1+\mathbf{K}_3=\mathbf{K}_2+\mathbf{K}_4$, where $\mathbf{K}_3$ and $\mathbf{K}_4$ represent the before and after wave-vectors for the \rC-region atoms participating in the collision. By definition the $\rC$-region atoms  must satisfy
 \begin{eqnarray}\label{Erestrict}
\frac{\hbar^2\mathbf{K}_j^2}{2m}+ V_{\rm eff}(\mathbf{R})\leq \ec,\quad j=3,4.
\end{eqnarray}
Since the momentum transfer in the scattering event satisfies $\hbar\mathbf{k}=\hbar\mathbf{K}_1-\hbar\mathbf{K}_2=\hbar\mathbf{K}_4-\hbar\mathbf{K}_3$, we have
\begin{eqnarray}
\frac{\hbar^2|\mathbf{k}|^2}{8m}+V_{\rm eff}(\mathbf{R})&=&\frac{1}{4}\left(\frac{\hbar^2\mathbf{K}_3^2}{2m}+\frac{\hbar^2\mathbf{K}_4^2}{2m}-\frac{\hbar^2\mathbf{K}_3\cdot\mathbf{K}_4}{m}\right) \nonumber \\ 
&&+V_{\rm eff}(\mathbf{R})\\
&\leq&\frac{1}{4}\left(\frac{\hbar^2\mathbf{K}_3^2}{2m}+\frac{\hbar^2\mathbf{K}_4^2}{2m}+\frac{\hbar^2|\mathbf{K}_3||\mathbf{K}_4|}{m}\right) \nonumber \\
&&+V_{\rm eff}(\mathbf{R})\\
&\leq&\ec.
\end{eqnarray}
Thus  $E_{\rm min}(\mathbf{R})\equiv \ec$ (in Ref.~\cite{SGPEII} this was stated as an approximation), and the $\mathbf{R}$ dependence is lost in Eq.~(\ref{Mgard}), i.e.~$M(\mathbf{R},\x)\to M(\x)$, which is the form we use  in this work.

\section{Outline of the numerical algorithm for the scattering SPGPE}\label{appa}

 Here we detail how we efficiently implement the scattering SPGPE in the harmonic oscillator basis.  The numerical implementation of the simple growth SPGPE has been previously outlined \cite{Bradley08a,Blakie08a}, being only slightly more complicated than integrating the PGPE for a harmonically trapped system \cite{Blakie08b}.  Thus we  focus here on our evaluation of the deterministic scattering effective potential and the associated scattering noise. We emphasize that here we present a simple overview of the method and a full and detailed account of our algorithm, particularly the use of quadratures to accurately and efficiently evaluate the necessary matrix elements, will be given elsewhere.

\subsection{Basis state representation}

We use a spectral representation of the c-field 
\eq{cfsumdmls}
{ \cf(\x,t)  = \sum_{n\in{\rC}} c_n(t) \phi_n( {\x}),}
in terms of the basis of harmonic oscillator modes $ \phi_n$ of the single-particle Hamiltonian satisfying $ H_{\rm sp} \phi_n =  \epsilon_n \phi_n$, where $c_n$ are time dependent complex amplitudes and $n$ represents all quantum numbers required to specify a single-particle state.
This choice is convenient because it allows us to efficiently implement the projection by restricting the spectral modes [as indicated in Eq.~(\ref{cfsumdmls})] to the set 
\eq{sumrestrict}
{\rC = \left\{ n:\epsilon_n \leq\epsilon_{\rm cut} \right\} .}

Projecting the scattering SPGPE \eref{scattspgpe} onto the basis-set modes in the $\rC$-region we obtain a system of equations for the evolution of the amplitudes, i.e.
\begin{equation}
(S)\,\,dc_n=-i[\epsilon_nc_n+G_n+S_n] dt +\sum_mB_{nm}\,dw_m,\label{dcn} 
\end{equation}

where
\begin{align}
G_n&\equiv  u\int d^3\x\,\phi_n^*(\x)\,|\psi (\x,t)|^2\psi (\x,t),\label{Gn} \\
S_n&\equiv  \int d^3\x\,\phi_n^*(\x)V_{ {M}}(\x,t)\psi (\x,t), \\
B_{nm}&\equiv -i\int d^3\x\,\phi_n^*(\x)\psi (\x,t) \chi_m(\x)\label{Bnm}, 
\end{align}
where we introduce the functions $\chi_m(\x)$ later [see Eq.~(\ref{chin})] and  $dw_m$ is the standard real Wiener process satisfying  $\langle dw_n\rangle=0$, $\langle dw_m dw_n\rangle=\delta_{mn}dt$.

There are two main steps in solving this equation: (i) time-evolution to step this equation forward in time; and (ii) evaluating the non-linear matrix elements (\ref{Gn})-(\ref{Bnm}) at each time step. 

\subsection{Time-evolution\label{sec:Eulerevo}}
We employ the \textit{weak vector semi-implicit Euler algorithm}   \cite{SM,Milstein,Drummond1991,Werner1997} to evolve our stochastic equations forward in time. As this algorithm is extensively discussed in the literature we just briefly review the algorithm here. Equation (\ref{dcn}) is of the general form \begin{equation}
(S)\,dc_n=a_n(t,\mathbf{c})\,dt+\sum_jB_{nm}(t,\mathbf{c})dw_{m}(t),
\end{equation}
where $a_n=-i[\epsilon_n c_n+G_n+S_n]$, and we use the notation $\mathbf{c}$ to represent the dependence of matrix elements on the full field ($\psi$). 
The solution is propagated to a set of discrete times $t_{j}=j\,\Delta t$, where $\Delta t$ is the step size, and we denote that solution at time  $t_j$ as $c^{(j)}_n$.  
Using this solution, the solution at the next time-step is computed as $c_n^{(j+1)}= c^{(j)}+\Delta c^{(j)}$, where
\begin{align}
 \Delta c^{(j)}_n&=a_n(\bar{t}_j,\bar{\mathbf{c}}^{(j)})\,\Delta t +\sum_mB_{nm}(\bar{t}_j,\bar{\mathbf{c}}^{(j)})\,\Delta w_m^{(j)},\label{SIE}
\end{align}
with 
\begin{align}
\bar{c}^{(j)}_n&\equiv\frac{1}{2}(c^{(j)}_n+c^{(j+1)}_n),\label{cbar}\\
\bar{t}_j&\equiv\frac{1}{2}(t_{j+1}+t_{j}),  \\
\langle \Delta w^{(j)}_m\Delta w^{(j)}_n\rangle &=\Delta t\, \delta_{mn}.\label{Deltaw}
\end{align}
Note formally $\Delta w^{(j)}_n\equiv\int_{t_j}^{t_{j+1}}dw_n$, however in practice we sample $\Delta w^{(j)}_m$ as a real Gaussian distributed random variable of variance $\Delta t$ [c.f.~Eq.~(\ref{Deltaw})].

Because $\bar{c}^{(j)}_n$ in Eq.~(\ref{SIE}) depends on $ c^{(j+1)}_n$ [i.e.~(\ref{cbar})], this equation is implicit. In practice a few iterations of Eq.~(\ref{SIE}) are usually sufficient to obtain convergence at small step-sizes.

This algorithm is correct to $O(\Delta t^2)$ in the limit of zero stochastic noise, and is convergent for the stochastic problem with a strong order of $\Delta t^{1/2}$ (i.e.~for individual trajectories) and with a weak order of $\Delta t^1$ (i.e.~for quantities calculated in the distribution).  

\subsection{Matrix elements}
As noted above the matrix elements of the usual Gross-Pitaevskii evolution [i.e.~(\ref{Gn})] and stochastic growth  are dealt with elsewhere \cite{Blakie08a,Blakie08b,Bradley08a} and we do not repeat this here. Instead we focus on the two new terms associated with scattering.

\subsubsection{Use of Fourier transforms to simplify scattering terms} 
In what follows we will use the notation
\begin{equation}
\tilde{f}(\mathbf{k})=\mathcal{F}\{f( \mathbf{r})\}\equiv\int\frac{d^3\mathbf{r}}{(2\pi)^{3/2}}e^{-i\mathbf{k}\cdot\mathbf{r}}f(\mathbf{r}),
\end{equation}
to denote the three-dimensional  Fourier transform of the function $f(\mathbf{r})$, which could be either a scalar or vector function, with associated inverse transform $f(\mathbf{r})=\mathcal{F}^{-1}\{\tilde{f}(\mathbf{k})\}$. 

These transforms can be efficiently and accurately implemented in the basis-set approach using the fact that the oscillator basis is the eigenbasis of the Fourier transform operator.  We note that the Fourier transform of the field (e.g.~$\mathcal{F}\{\psi\}$) and density-type quantities (e.g.~$\mathcal{F}\{|\psi|^2\}$) are treated in different ways, as discussed in Ref.~\cite{Blakie2009a}.

\subsubsection{Scattering effective potential} 
Using (\ref{Mdef}) 
and the convolution theorem, the scattering effective potential can be evaluated as
\eq{Vmk}{ 
V_M(\x)=-\frac{\hbar^2{\cal M}}{k_BT}\mathcal{F}^{-1}\left\{i\hat{\kk}\cdot\mathcal{F}\{{\mathbf{j}}(\x)\}\right\},
}
where $\hat{\kk}=\kk/|\kk|$. The current $\mathbf{j}(\x)$ is quite conveniently formed in the oscillator basis making use of the step operators to take the spatial derivatives.

Equation (\ref{Vmk}) reveals that the scattering effective potential depends on the radial part of the current in $\kk$-space. This corresponds to the irrotational part of the  of the current, i.e.
\eq{Vmk2}{ 
V_M(\x)=-\frac{\hbar^2\mathcal{M}}{k_BT}\bm{\nabla}\cdot\mathbf{j}_{\parallel},
}
where $\mathbf{j}(\x)=\mathbf{j}_{\parallel}(\x)+\mathbf{j}_{\perp}(\x)$ is the Helmholtz decomposition with  $\nabla\times\mathbf{j}_{||}(\x)=\nabla\cdot\mathbf{j}_{\perp}(\x)=0$.

\subsubsection{Scattering noise} 
While the scattering noise \eref{dwm1} has a non-local correlation function in position space, in Fourier space it satisfies
\eqn{\label{dwk2}
\langle d\tilde{W}_{M}(\kk)d\tilde{W}_{M}(\kk^\prime)\rangle&=&\frac{2{\cal M}dt}{(2\pi)^3}\mint{\x}e^{-i\kk\cdot\x}\mint{\x^\prime}e^{-i\kk^\prime\cdot\x^\prime} \nonumber \\
&\times&\mint{\q}\frac{e^{i\q\cdot(\x-\x^\prime)}}{(2\pi)^3|\q|}, \\
&=&\frac{2{\cal M}dt}{|\kk|}\delta(\kk+\kk^\prime).\label{targetcorreln}
}
That is, the noise is  \emph{anti-diagonal} in $\kk$-space. We implement the noise generation procedure  using the following result:

Choosing the oscillator basis to be separated into products of one-dimensional oscillator eigenstates $\phi_n(\x)=\varphi^{(x)}_{n_x}(x)\varphi^{(y)}_{n_y}(y)\varphi^{(z)}_{n_z}(z)$, where we have decomposed the quantum number as $n=\{n_x,n_y,n_z\}$. These 1D states are even or odd, as determined by their quantum number, e.g.~$\varphi^{(x)}_{n_x}(x)=(-1)^{n_x}\varphi^{(x)}_{n_x}(-x)$, and thus full basis states have the parity property
\begin{equation}
\phi_n(\x)=(-1)^\sigma\phi_n(-\x),\label{Phiparity}
\end{equation}
where  $\sigma=n_x+n_y+n_z$.
Taking the oscillator basis to be real (in position space), the Fourier transformed basis modes $\tilde{\phi}_n(\kk)$ are then purely real or imaginary functions depending on this symmetry, i.e.
\begin{equation} 
\tilde{\phi}_n(\kk)=(-i)^\sigma{\Phi}_n(\kk),
\end{equation}
where  the ${\Phi}_n(\kk)$ are a purely-real set of orthonormal orbitals. Importantly, the functions ${\Phi}_n(\kk)$ are, to within a scaling along each dimension, identical to the position space functions $\phi_n(\x)$, and thus property (\ref{Phiparity}) holds
\begin{equation}
\Phi_n(\kk)=(-1)^\sigma\Phi_n(-\kk).
\end{equation}

Using this result, the scattering noise is then constructed in $\kk$-space as  
\begin{equation}
d\tilde{W}(\kk)=\sqrt{\frac{2\mathcal{M}}{ |\kk|}}\sum_n\tilde{\phi}_n(\kk)\, dw_n,
\end{equation}
which has the desired  correlation function  (\ref{targetcorreln}):
\begin{align}
\left\langle d\tilde{W}(\kk)d\tilde{W}(\kk') \right\rangle &=\frac{2\mathcal{M}dt}{|\kk|}\sum_n(-1)^\sigma\Phi_n(\kk)\Phi_n(\kk'),\\
& =\frac{2\mathcal{M}dt}{|\kk|}\sum_n \Phi_n(\kk)\Phi_n(-\kk'),\\
& =\frac{2\mathcal{M}dt}{|\kk|}\delta_{\rC}(\kk,-\kk'),
\end{align}
where
$\delta_{\rC}(\kk,\kk')=\sum_n\Phi_n(\kk)\Phi_n(\kk')=\sum_n\phi_n^*(\kk)\phi_n(\kk')$ is the projected delta function.

Thus, in $\x$-space the noise is given by
\begin{equation}
dW_M(\x,t) = \sum_m\chi_m(\x)\,dw_m,
\end{equation}
where we have introduced 
\begin{equation}
\chi_m(\x)\equiv\mathcal{F}\left\{\sqrt{\frac{2\mathcal{M}}{ |\kk|}}\tilde{\phi}_m(\kk)\right\} \label{chin}.
\end{equation}



\end{document}